\documentclass[aps,prb,twocolumn,superscriptaddress]{revtex4-1}
\usepackage{graphicx}
\begin{document}


\title{Transmission Enhancement of High-$k$ Waves through Metal-InGaAsP Multilayers	Calculated via Scattering Matrix Method with Semi-Classical Optical Gain}


\author{Joseph S. T. Smalley}
\email[]{jsmalley@ucsd.edu}
\altaffiliation{}
\affiliation{Department of Electrical and Computer Engineering, University of California San Diego, La Jolla, CA 92103}
\affiliation{Center for Integrated Nanotechnologies, Sandia National Laboratory, Albequerque, NM 87185}
\author{Felipe Vallini}
\affiliation{Department of Electrical and Computer Engineering, University of California San Diego, La Jolla, CA 92103}
\author{Shiva Shahin}
\affiliation{Department of Electrical and Computer Engineering, University of California San Diego, La Jolla, CA 92103}
\author{Boubacar Kant\'{e}}
\affiliation{Department of Electrical and Computer Engineering, University of California San Diego, La Jolla, CA 92103}
\author{Yeshaiahu Fainman}
\affiliation{Department of Electrical and Computer Engineering, University of California San Diego, La Jolla, CA 92103}

\date{\today}

\begin{abstract}
We analyze the steady-state transmission of high-momentum (high-$k$) electromagnetic waves through metal-semiconductor multilayer systems with loss and gain in the near-infrared (NIR).  Using a semi-classical optical gain model in conjunction with the scattering matrix method (SMM), we study indium gallium arsenide phosphide (InGaAsP) quantum wells as the active semiconductor, in combination with the metals, aluminum-doped zinc oxide (AZO) and silver (Ag).  Under moderate external pumping levels, we find that NIR transmission through Ag/InGaAsP systems may be enhanced by several orders of magnitude relative to the unpumped case, over a large angular and frequency bandwidth.  Conversely, transmission enhancement through AZO/InGaAsP systems is orders of magnitude smaller, and has a strong frequency dependence.  We discuss the relative importance of Purcell enhancement on our results and validate analytical calculations based on the SMM with numerical finite-difference time domain simulations.   
\end{abstract}


\maketitle

\section{\label{Section1}Introduction}
Metal-dielectric (MD) interfaces support optical surface waves whose effective wavelength is less than the smallest wavelength achievable in dielectrics\cite{RefWorks:212}.  The coupling of multiple MD interfaces enables even smaller effective wavelengths, or conversely higher effective indices, or so-called high-$k$ propagation, of the optical waves\cite{RefWorks:183,RefWorks:209,RefWorks:190}.  The phenomenon of large effective indices and propagating high-$k$ modes lies at the heart of sub-diffraction-limited imaging devices\cite{RefWorks:113,RefWorks:169,RefWorks:74}, sensors based on plasmonic resonances\cite{RefWorks:219,RefWorks:216}, as well as applications of hyperbolic metamaterials (HMMs), including asymmetric transmission devices\cite{RefWorks:191}, nonlinear optics\cite{RefWorks:218}, and lifetime reduction of dye molecules\cite{RefWorks:67,RefWorks:193} and quantum dots\cite{RefWorks:199}.  Inherent to the increase in the effective index in MD systems, is a concomitant increase in dissipation losses\cite{RefWorks:189,RefWorks:217}.  The tradeoff between optical confinement and losses is a hallmark of plasmonic and hyperbolic media, and may be considered one of the greatest challenges to the widespread realization of integrated plasmonic technology\cite{RefWorks:222,RefWorks:102}. 

Strategies for reducing losses in plasmonic and HMMs have therefore been proposed, including clever design of passive structures and incorporation of active, i.e. gain media\cite{RefWorks:80,RefWorks:70,RefWorks:3,RefWorks:69}.  An exemplary case of the former strategy is the fishnet metamaterial\cite{RefWorks:214,RefWorks:215}, whereby air voids in an otherwise continuous MD multilayer open new channels for optical transmission.  Even in these structures though, gain media are still necessary to obtain low-loss transmission at optical frequencies, as was demonstrated via infiltrated optical dyes in the core of the structure\cite{RefWorks:109}.      

While dye molecules are straightforward to model as non-interacting two-level systems and convenient for proof-of-concept experiments\cite{RefWorks:109,RefWorks:224,RefWorks:223,RefWorks:226}, an ideal gain media in plasmonic systems would be electronically addressable\cite{RefWorks:234}.  In this case, semiconductor heterostructures become strong candidates for incorporation into plasmonic systems.  III-V semiconductors emitting in the near-infrared (NIR) have been studied in conjunction with surface plasmon amplification\cite{RefWorks:227}, and successfully implemented in sub-wavelength metal-coated and plasmonic sources\cite{RefWorks:227,RefWorks:182,RefWorks:147}.  However, the incorporation of active III-V materials in multilayer MD systems with hyperbolic dispersion remains to be demonstrated in the NIR.    

\begin{figure}
\includegraphics[width=8.5cm]{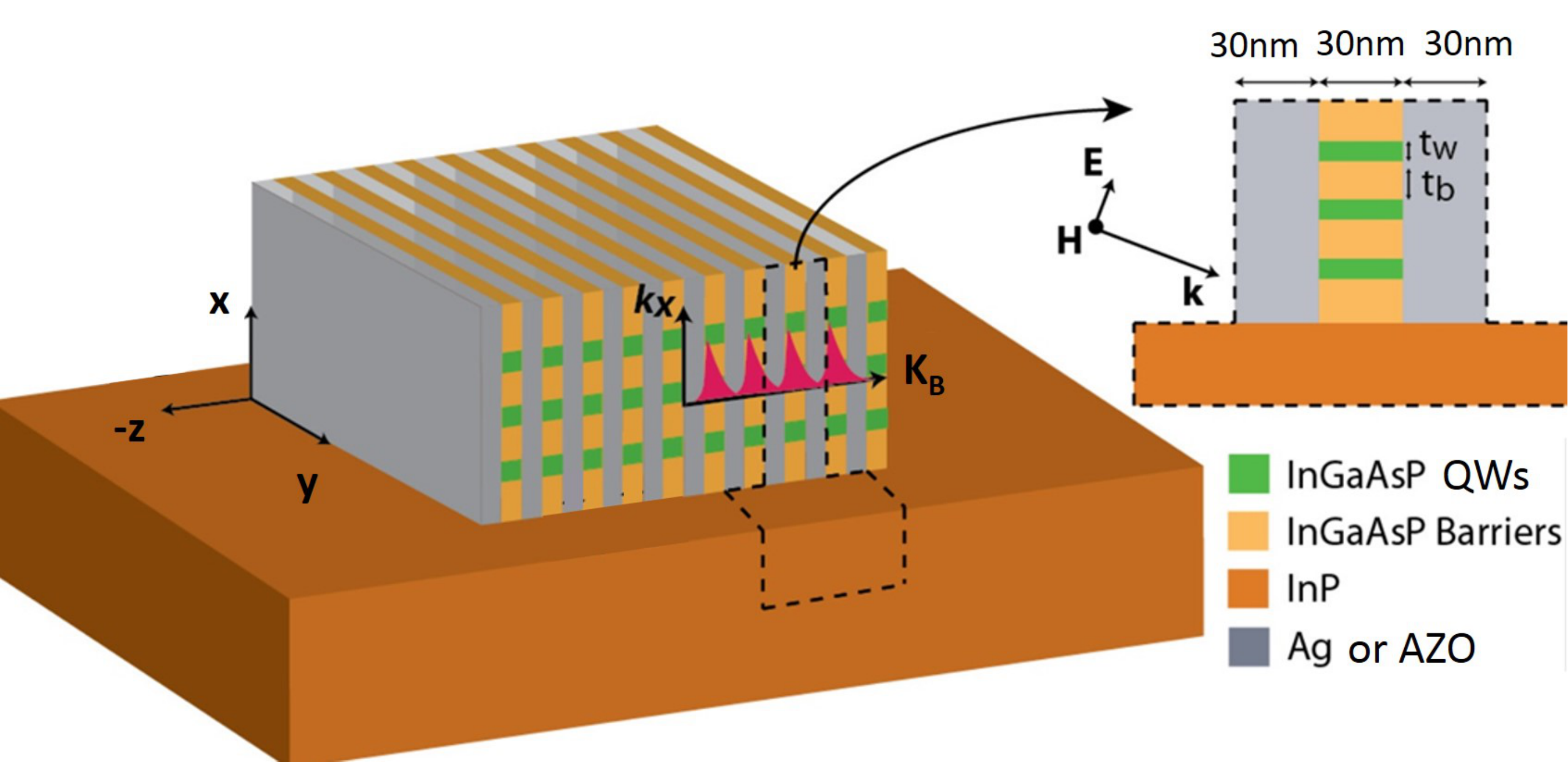}%
\caption{\label{Fig1}Schematic of multilayer metal-InGaAsP MQW system.  The heterostructure, with $t_{well}$=10nm and $t_{barrier}$=20nm is grown in the y-direction and TM-polarized light propagates in the z-direction.}
\end{figure}

In this work, we apply the scattering matrix method\cite{RefWorks:141} (SMM), to analyze the possibility for using indium gallium arsenide phosphide (InGaAsP) multiple quantum wells (MQWs) as a gain media in NIR MD systems.  InGaAsP is a mature quaternary III-V compound that is epitaxially grown on indium phosphide (InP) substrates.  This immediately poses challenges for its incorporation in a multilayer MD system.  We envision deposition of metallic thin films into finely patterned trenches between InGaAsP pillars formed by nanoimprint lithography and subsequent reactive ion etching.  We assume that the smallest layer thicknesses enabled by this process are 30nm.  The resulting MD system, shown schematically in Fig.\ref{Fig1}, supports volume plasmon polaritons\cite{RefWorks:190} (VPPs) with large effective indices and large loss in the absence of external pumping, the hallmark of strongly coupled MD interfaces.  These high-$k$ VPPs propagate normal to the epitaxial growth direction, which is a novel multilayer configuration, potentially more suitable to waveguide-integrated HMMs, compared to the more common large-area multilayers which are conformal to the wafer substrate.   In the following, we show that under moderate external pumping conditions, MD systems composed of InGaAsP MQWs may support transmission of high-index modes enhanced by several orders of magnitude, relative to the unpumped case.

In Section \ref{Section2}, we introduce the optical gain model employed, in conjunction with the SMM.  We present the main results of the report in Section \ref{Section3}, namely, the NIR transmission characteristics of multilayer MD systems based on InGaAsP MQWs.  We investigate InGaAsP in combination with aluminum-doped zinc oxide (AZO), which is considered a low-loss transparent conducting oxide and alternative plasmonic material for NIR applications\cite{RefWorks:72,RefWorks:71,RefWorks:108}, and in combination with silver, which is well known as the noble metal with highest conductivity\cite{RefWorks:115}.  In Section \ref{Section4}, we discuss the relative importance of Purcell enhancement on our results.  We also validate the analytical SMM results with numerical finite-difference time-domain (FDTD) simulations.  We then discuss the limitations of our results imposed by the assumptions of our models.  We conclude the report in Section \ref{Section5}.

\section{\label{Section2}Methods}

The quaternary III-V compound that we consider is an In$_x$Ga$_{1-x}$As$_y$P$_{1-y}$ MQW system where (x=0.564, y=0.933) and (x=0.737, y=0.569) for the wells and barriers, respectively.  The nominal room-temperature bandgap energies of the barrier and well materials are 0.953eV ($\lambda_G$=1.3$\mu$m) and 0.774eV ($\lambda_G$=1.6$\mu$m), respectively. This system can be epitaxially grown on an InP substrate and has been experimentally utilized in near-infrared subwavelength semiconductor lasers\cite{RefWorks:147,RefWorks:160}.  

In the dipole-approximation the peak magnitude of the optical gain is governed by the transition matrix element, $M_T$.  Because MD multilayers only support propagation of TM-polarized modes\cite{RefWorks:209}, we consider TM-polarized light emission from the quantum wells.  To justify the use of a steady-state analysis and the non-inclusion of dynamic nonlinear effects, and for simplicity, we concern ourselves with moderate pumping levels, about two times above the transparency inversion density.  Consequently, we may assume that most of the transitions occur at the band-edge such that recombination between electrons and holes with vanishing transverse momentum dominates\cite{RefWorks:44}.  In this limit, only transitions from the conduction band to the light-hole band contribute to the matrix element for TM-polarization\cite{RefWorks:44}.  We approximate the bandstructure of InGaAsP with parabolic conduction and valence bands, characterized by the conduction and light-hole effective masses of 0.0481$m_0$ and 0.0537$m_0$, respectively\cite{RefWorks:210}, where $m_0$=9.1x10$^{-31}$1kg is the free electron mass.  The magnitude of the matrix element used is given by\cite{RefWorks:220,RefWorks:221} $|M_T|^2$=(24.9eV)$m_0$/3, which corresponds to a dipole length of $|$x$|\approx$2.8nm at $\lambda_0$=1500nm.  

Employing this gain model we obtain complex permittivity values for a 10nm InGaAsP QW at carrier densities representative of absorbing (Abs) and moderately inverted (Inv) states, $N$=1.0x16cm$^{-3}$ and $N$=5.0x18cm$^{-3}$, respectively.  At $\lambda_0$=1500nm, the imaginary parts of these values are $\epsilon_{D,Abs}"$=$\Gamma$g($N$=1.0x16cm$^{-3}$)=+0.139 and $\epsilon_{D,Inv}"$=$\Gamma$g($N$=5.0x18cm$^{-3}$)=-0.143, where $\Gamma$=1/3 is the relative area of wells in the InGaAsP heterostructure. Using Kramers-Kronig relations\cite{RefWorks:235}, the real parts at $\lambda_0$=1500nm are $\epsilon_{D,Abs}'$=11.997 and $\epsilon_{D,Inv}'$=11.881, respectively.  For the sake of simplicity, we used a carrier-independent but frequency dependent real permittivity, which is  $\epsilon_D'$=11.914 at $\lambda_0$=1500nm, and justified because the transmission properties studied are dominated by loss and gain.  Relating the imaginary permittivity to the linear gain coefficient\cite{RefWorks:227}, g=-$k_0$$\epsilon_D"$$(\epsilon_D')^{-1/2}$, the value of $\epsilon_D"=\pm$0.14 corresponds to a loss/gain per unit length of $\pm$1700cm$^{-1}$  at 1500nm. The transparency (Tra) condition\cite{RefWorks:44}, $\epsilon_{D,Tra}"$=0, for InGaAsP at $\lambda_0$=1500nm corresponds to about $N$=2.0x18cm$^{-3}$.  The complex permittivity values for AZO\cite{RefWorks:71} and Ag\cite{RefWorks:38} at $\lambda_0$=1500nm are -0.392+i0.139 and -122.190+i3.115, respectively.  Additional details on the gain model may be found in the references\cite{RefWorks:44,RefWorks:104}.  

Combined with the presented optical gain model, we use the SMM as implemented by Krayzel et al.\cite{RefWorks:207} to study the transmission properties of an idealization of the structure of Fig.\ref{Fig1}.  In comparison to simpler methods such as effective medium theory\cite{RefWorks:141}, Bloch\textquoteright s theorem\cite{RefWorks:69}, and the transfer matrix method\cite{RefWorks:70,RefWorks:197,RefWorks:3}, the SMM is well known as a superior method for analyzing systems with evanescent waves\cite{RefWorks:204,RefWorks:201} and strong loss/gain\cite{RefWorks:200}.  The idealized structure includes loss and gain, is infinite in the transverse (x-y) plane and surrounded by a uniform medium in the longitudinal (z) direction.  To couple to the high-$k$ modes supported by the MD multilayer of Fig.\ref{Fig1}, the surrounding medium takes the form of a prism.  The prism behaves as a numerical simplification to the physically equivalent grating-coupling technique\cite{RefWorks:191,RefWorks:67,RefWorks:199,RefWorks:213,RefWorks:206}, which is required for practical excitation of the high-$k$ modes supported by each system. 

\begin{figure}
	\includegraphics[width=9cm]{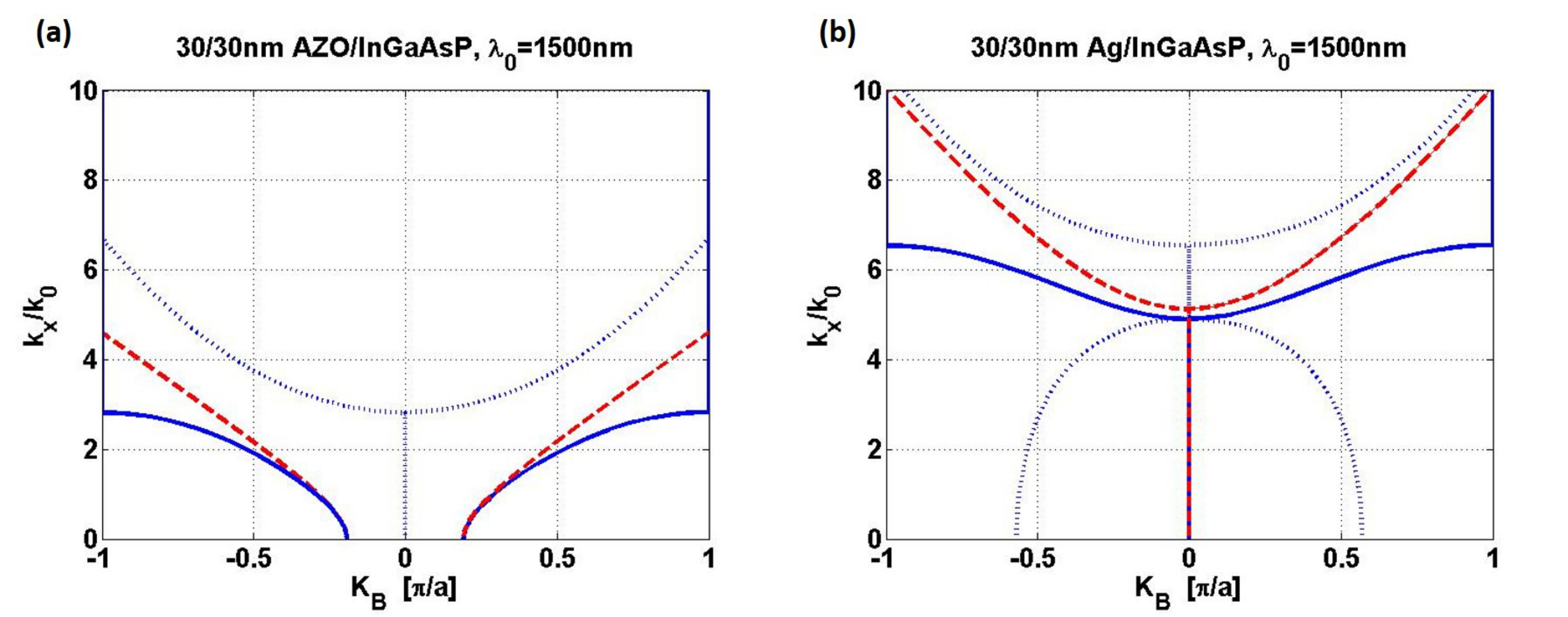}%
	\caption{\label{Fig2}Wave-vector diagrams for 30/30nm (a) AZO/InGaAsP and (b) Ag/InGaAsP systems at 1500nm with losses omitted.  (solid blue line=real part of Bloch vector, dotted blue line=imaginary part of Bloch vector, dashed red line=effective medium theory prediction of real part of Bloch vector)  The transmission windows in (a) and (b) extend from 0$\leq$$k_x$$\leq$2.83$k_0$ and 4.91$k_0$$\leq$$k_x$$\leq$6.56$k_0$, respectively.}
\end{figure}

To determine a suitable value of the prism permittivity, $\epsilon_P$, we first determine the bounds of the high-$k$ transmission windows in each MD system.  This is achieved using  Bloch\textquoteright s theorem\cite{RefWorks:188} in the absence of losses by locating the regions in momentum space wherein the real part of the Bloch vector is purely real. For MD systems with 30nm layer thicknesses and $\lambda_0$=1500nm, Fig.\ref{Fig2}(a) and \ref{Fig2}(b) show the real (solid blue curve) and imaginary (dotted blue curve) parts of the Bloch vector, indicating that transmission windows exist when 0$\leq$$k_x$$\leq$2.83$k_0$ and 4.91$k_0$$\leq$$k_x$$\leq$6.56$k_0$ in AZO/InGaAsP and Ag/InGaAsP systems, respectively.  In the limit of effective medium theory (dashed red curves), Fig. \ref{Fig2}(a) and \ref{Fig2}(b) are representative of type-I and type-II HMMs, respectively.  Based on these high-$k$ windows, we set $\epsilon_P$=64 for the SMM calculations.     

\section{\label{Section3}Results}

\begin{figure}
	\includegraphics[width=9cm]{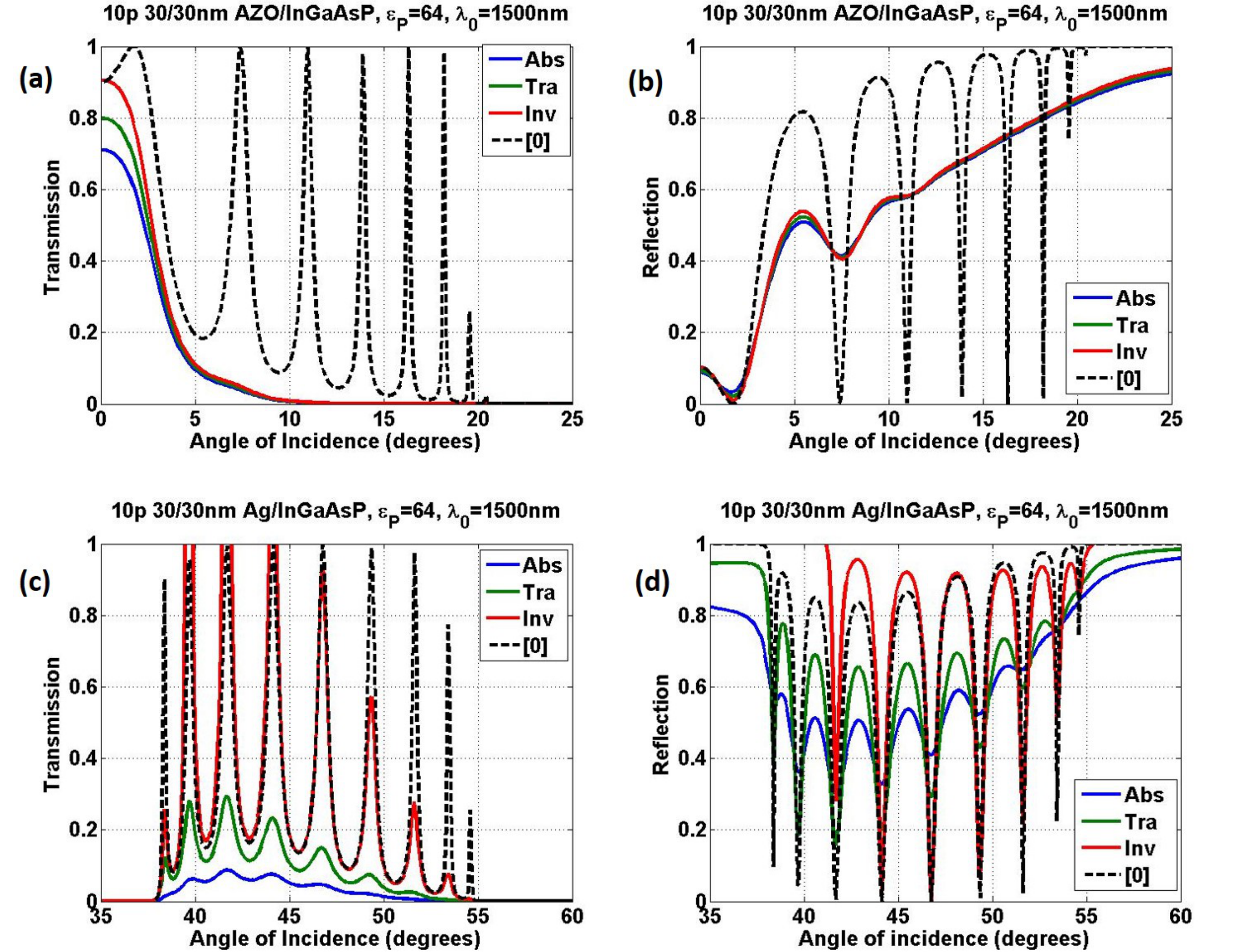}
	\caption{\label{Fig3}(a,c) Transmission and (b,d) reflection for TM-polarized light of wavelength 1500nm, incident on 10-period (a,b) AZO/InGaAsP and (c,d) Ag/InGaAsP multilayer with 30nm layers, coupled via prism with $\epsilon_P$=64. (Abs: $N$=1x10$^{16}$cm$^{-3}$; Tra: $\epsilon_D"$=0; Inv: $N$=5x10$^{18}$cm$^{-3}$; [0]: $\epsilon_M"$=$\epsilon_D"$=0)}
\end{figure}

Transmission and reflection at $\lambda_0$=1500nm through a 10-period AZO/InGaAsP structure with 30nm layer thicknesses, is shown as a function of incident angle in Fig.\ref{Fig3}(a) and \ref{Fig3}(b), respectively.  The resonances extending from normal incidence to $\theta_{inc}$=20$^{\circ}$ translate into a high-$k$ window that extends to roughly $k_x$=8$\sin$(20$^{\circ}$)$\approx$2.8$k_0$, consistent with predictions based on Bloch\textquoteright s theorem (Fig. \ref{Fig2}(a)).  The effect of inverting the carrier population in absolute terms appears most pronounced near normal incidence and becomes weaker as the incident angle increases.  The strong absorption in this system is apparent in the local reflection minimum around $\theta_{inc}$=7.5$^{\circ}$.  The lack of corresponding transmission maximum indicates strong dissipation, even when the semiconductor is inverted.  

Transmission and reflection through the AZO/InGaAsP system is best appreciated when contrasted to the behavior of the Ag/InGaAsP system.  In Fig.\ref{Fig3}(c) and \ref{Fig3}(d), transmission and reflection for the latter system are shown, with identical conditions to the AZO-based structure.  The transmission window spans approximately 5.0$k_0$$\leq$$k_x$$\leq$6.5$k_0$, again consistent with Bloch\textquoteright s theorem (Fig. \ref{Fig2}(b)).  The effect of inverting the carrier population becomes dramatic in both absolute and relative terms.  For several resonances between 5.0$k_0$ and 5.7$k_0$, the transmission and reflection exceed unity, indicating active behavior of the system.

\begin{figure}
	\includegraphics[width=9cm]{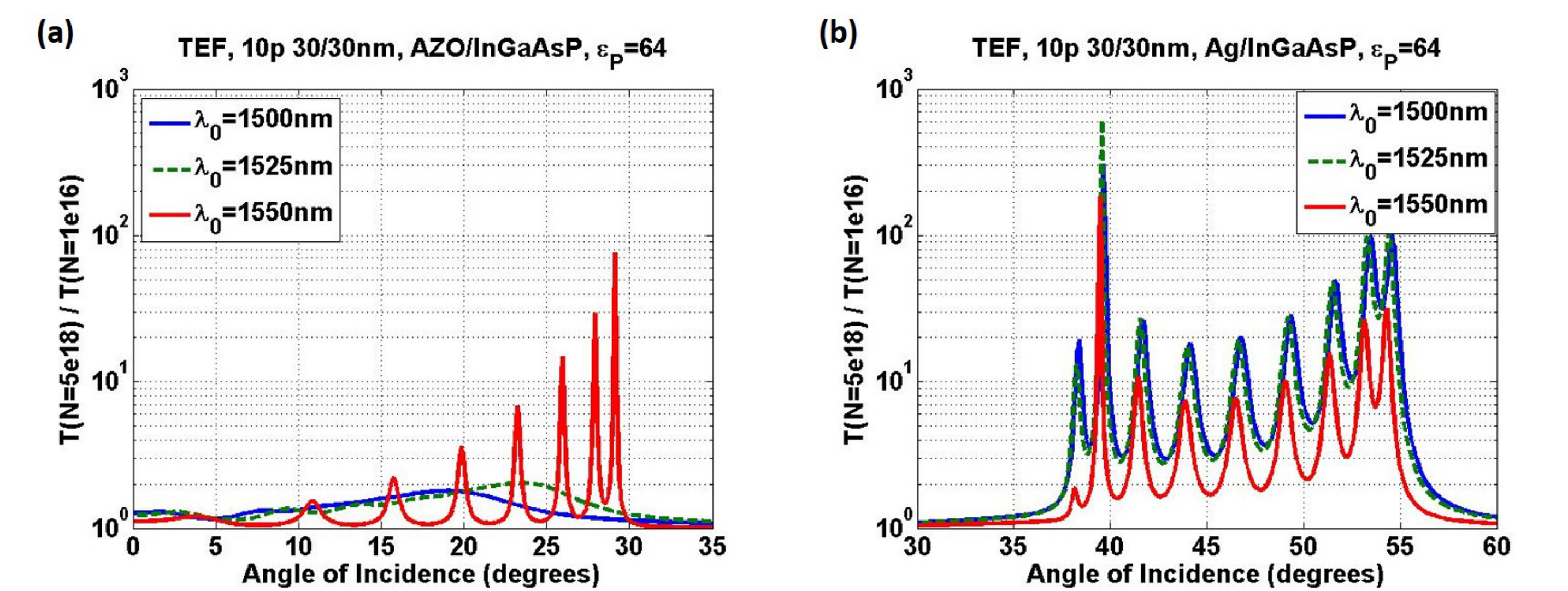}%
	\caption{\label{Fig4}Transmission enhancement factor ($TEF$) for TM-polarized light incident on 10-period (a) AZO/InGaAsP and (b) Ag/InGaAsP multilayer with 30nm layers, coupled via prism with $\epsilon_P$=64.}
\end{figure}

To quantify the effect of the inverted carrier population on the behavior of the MD systems, we define the relative transmission enhancement factor ($TEF$) as
\begin{equation}
TEF\equiv T(5\textrm{x}10^{18}\textrm{cm}^{-3})/T(1\textrm{x}10^{16}\textrm{cm}^{-3}),
\end{equation}
the transmission under inversion with respect to transmission under absorption of the InGaAsP MQW.   Fig.\ref{Fig4}(a) shows the $TEF$ for the AZO-based multilayer at several different incident wavelengths.  This system exhibits significant dispersion due to the proximity of the plasma frequency of AZO to the NIR.  For shorter wavelengths, enhancement is negligible, indicating that realistic gain levels cannot compete with the strong damping at the plasmon resonance.  As the wavelength increases to 1550nm, a modest $TEF$ appears, increasing with the incident angle.  However, this relative enhancement must be celebrated cautiously.  The improvement in the $TEF$ with increasing angle of incidence is offset by a drop in the absolute transmission (Fig.\ref{Fig3}(a)).  Consequently, the overall effect of carrier inversion is quite small in the AZO/InGaAsP system.

Again, these results are best appreciated by contrasting to the Ag/InGaAsP system, the $TEF$ of which is shown in Fig.\ref{Fig4}(b).  Relative enhancements close to a factor of 10 are observed over the range of incident angles supported by this system, with several prominent peaks exceeding factors of 100 near the edges of the transmission window.  In comparison to the AZO/InGaAsP system, the $TEF$ has weak dispersion and is consistently larger in magnitude, indicating that the effect of the inversion in the Ag/InGaAsP system is much stronger.  Thus, if efficient coupling to this system is achieved, tunable and/or active, pump-dependent behavior may be feasible, enabling, for example, extremely compact and electronically addressable optical amplifiers, sources, and mixers.

\section{\label{Section4}Discussion}
Prior work on gain-compensated plasmonic systems has emphasized that the Purcell effect has a deleterious role\cite{RefWorks:234,RefWorks:197}.  Specifically, while the carrier densities required for significant improvement in transmission appear practically achievable, the required current densities may be extremely high\cite{RefWorks:234}.  This is a consequence of the large optical density of states in plasmonic and HMMs and the related high rates of spontaneous emission\cite{RefWorks:96}.  We stress, however, that the Purcell effect in MD systems is most significant near the plasma frequency of the constituent metal and becomes less important as the frequency decreases.  Using the expression\cite{RefWorks:234} $F_P$=1+$\pi$$\Gamma$$k_{z,D}$$k_x$$\omega$($dk_x$/$d\omega$)/($\epsilon_D$$k_0$)$^3$ for the Purcell factor, $F_P$, at a single MD interface, where $k_{z,D}$ is the longitudinal wave-vector component in the dielectric, and $\omega$ and $k_0$ are the angular frequency and vacuum wave-number, respectively, we immediately notice that the Purcell enhancement in the AZO/InGaAsP system is significantly greater than unity due to the proximity of the AZO plasma frequency to the NIR (Appendix \ref{App1}).  On the other hand, the Purcell factor for the Ag/InGaAsP system is on the order of unity because $dk_x$/$d\omega$$\approx$0 (Appendix \ref{App1}).  Therefore, while Purcell enhancement may cause the already stringent gain requirements for the AZO/InGaAsP system to increase, they may be considered a minor concern for the Ag/InGaAsP system.

\begin{figure}
	\includegraphics[width=9cm]{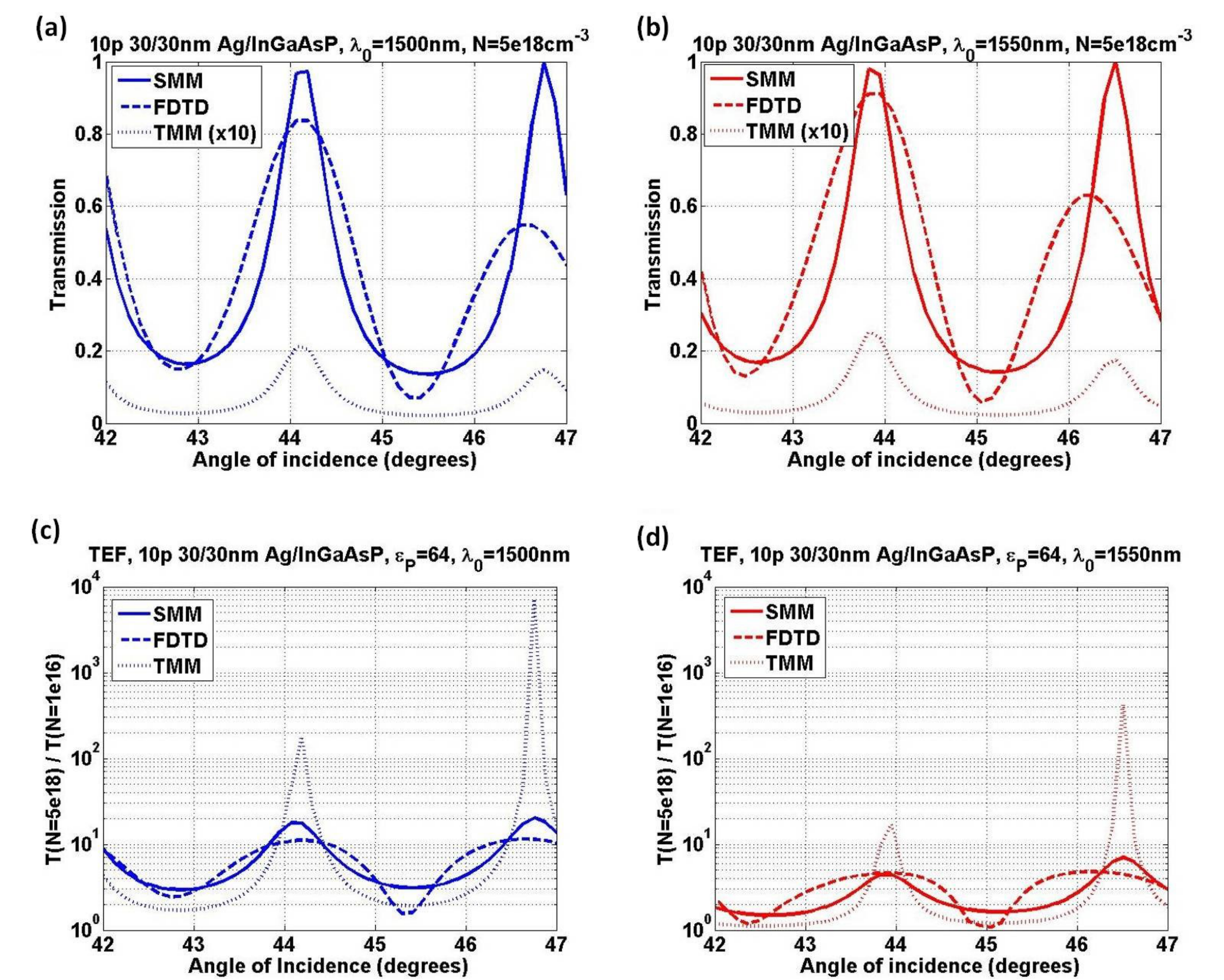}%
	\caption{\label{Fig5}(a,b) Transmission and (c,d) $TEF$ through prism-coupled 10-period 30/30nm Ag/InGaAsP system under inversion at $\lambda_0$= (a,c) 1500nm and (b,d) 1550nm.  The TMM result has been increased 10x in (a,b) for visual comparison. (SMM=scattering matrix method, FDTD=finite-difference time-domain, TMM=transfer matrix method)}
	
\end{figure}

To validate the SMM results on MD systems, we performed numerical FDTD simulations (Lumerical$\textregistered$) of the transmission through a 10-period Ag/InGaAsP system with 30nm layers under the prism coupling configuration and with material parameters identical to the SMM.  Periodic boundary conditions and perfectly matched layers (PML) were enforced in the transverse and longitudinal directions, respectively.  For each simulation, the center frequency of the finite-bandwidth pulse was tuned to match the monochromatic SMM calculations.  Figures \ref{Fig5}(a) and \ref{Fig5}(b) compare transmission through this system at 1500nm and 1550nm, respectively, over a narrow angular range.  The SMM is seen to consistently agree with FDTD results within a factor of two. Over the complete transmission window of the Ag/InGaAsP system, agreement between the SMM and FDTD worsens (Appendix \ref{App2}).  This discrepancy between the SMM and FDTD may be caused for several reasons, including imperfect PML and the finite bandwidth of the incident wave used in the FDTD simulations.  While the exact source of the discrepancy is unknown, we offer several possibilities in Appendix \ref{App2}.  Figure \ref{Fig5}(a-b) additionally includes results based on the TMM, which disagrees with predictions of SMM and FDTD by several orders of magnitude over the entire transmission window (further details in Appendix \ref{App1}). 

Throughout this analysis we have assumed a steady-state gain model, which implies continuous wave external pumping conditions.  A logical extension of the present work is a FDTD study incorporating spatial dependence of the gain within each semiconductor layer, as well as the use of a self-consistent gain model that takes into account transient and nonlinear effects\cite{RefWorks:155}.  The inhomogeneously broadened lineshape of InGaAsP MQWs however poses challenges in this regard, as a simple 4-level system\cite{RefWorks:150} does not capture the electronic density of states responsible for the broadening.  A potential solution to this problem is the use of a superposition of 4-level systems, each with a slightly different transition frequency.  The added accuracy of this approach, however, would come at the expense of large memory and processing requirements.  

\section{\label{Section5}Conclusion}
We have analyzed transmission of high-$k$ waves in the presence of loss and gain through metal-InGaAsP systems.  We have shown that, under moderate pumping, Ag/InGaAsP may support transmission of high-$k$ modes enhanced by several orders of magnitude, relative to the unpumped case, and absolute transmission on the order of unity.  While the magnitude of the enhancement factor for the AZO/InGaAsP system is significant, the absolute transmission is rather negligible.  We have further shown that the Purcell effect does not have a strong deleterious role for the Ag/InGaAsP system.  The results provide justification for experimental efforts on multilayer MD systems based on InGaAsP MQWs that circumvent the confinement-loss tradeoff fundamental to plasmonic and HMMs, as well as a rigorous foundation for further theoretical work on the nonlinear dynamics of such systems.

\acknowledgments
The authors wish to thank Prof. Antione Moreau for helpful discussions.  This work was supported by the Office of Naval Research Multidisciplinary Research Initiative (N00014-13-1-0678), the National Science Foundation (NSF) (ECE3972 and ECCS-1229677), the NSF Center for Integrated Access Networks (EEC-0812072, Sub 502629), the Defense Advanced Research Projects Agency (N66001-12-1-4205), and the Cymer Corporation. 

\bibliography{Smalley3}

\appendix
\label{Appendix}
\section{\label{App1}Use of Bloch's Theorem and the TMM in analysis of MD systems}
To date, most work on the analysis of MD systems and HMMs has relied upon the transfer matrix method (TMM).  For example, greater-than unity transmission through Ag/TiO$_2$ multilayers was predicted via TMM by Cortes et al\cite{RefWorks:96}.  This result is unphysical because loss and gain were omitted.  In the presence of loss and gain, Savelev et al. used the TMM method to analyze transmission through Ag/PMMA systems\cite{RefWorks:197}.  While the reported transmission is plausible, no attempt was made to validate the results with numerical methods.

In addition to transmission, the TMM is routinely used to calculate the complex amplitude reflectivity of HMMs, upon which the photonic density of states and emission lifetimes are often calculated.  This was done, for example, by Galfsky et al. on quantum dots in Ag/Al$_2$O$_3$ multilayers\cite{RefWorks:199}, by Lu et al. and Ferrari et al. on dyes in Ag/Si systems\cite{RefWorks:67,RefWorks:168}, and by Cortes et al. for Ag/TiO$_2$\cite{RefWorks:96}. The TMM was also used to elucidate the presence of volume plasmon polaritons in Au/Al$_2$O$_3$ multilayers by Zhukovsky et al.\cite{RefWorks:96}, and to calculate the bandstructure of HMMs in the presence of gain\cite{RefWorks:184}.

While useful qualitatively, the TMM is known to become increasingly numerically unstable as the structure increases in length\cite{RefWorks:204}, as strong loss or gain is incorporated\cite{RefWorks:200}, and as evanescent modes are considered\cite{RefWorks:202}.  The essential difference between the TMM and SMM lies in the fact that the matrices of the former contain exponential functions with arguments of different sign\cite{RefWorks:204,RefWorks:200,RefWorks:202,RefWorks:203,RefWorks:201}.  For large propagation lengths or large values of the imaginary component of the wave-vector, one of the exponentials will diverge.  The matrices of the SMM on the other hand contain exponentials with arguments of the same sign, which prevent the instabilities that limit the utility of the TMM.  Given these facts, it is imperative to use a more stable method for calculating absolute transmission of high-$k$ modes in the presence of loss and gain, as well as relative transmission enhancement.

\begin{figure}
	\includegraphics[width=9cm]{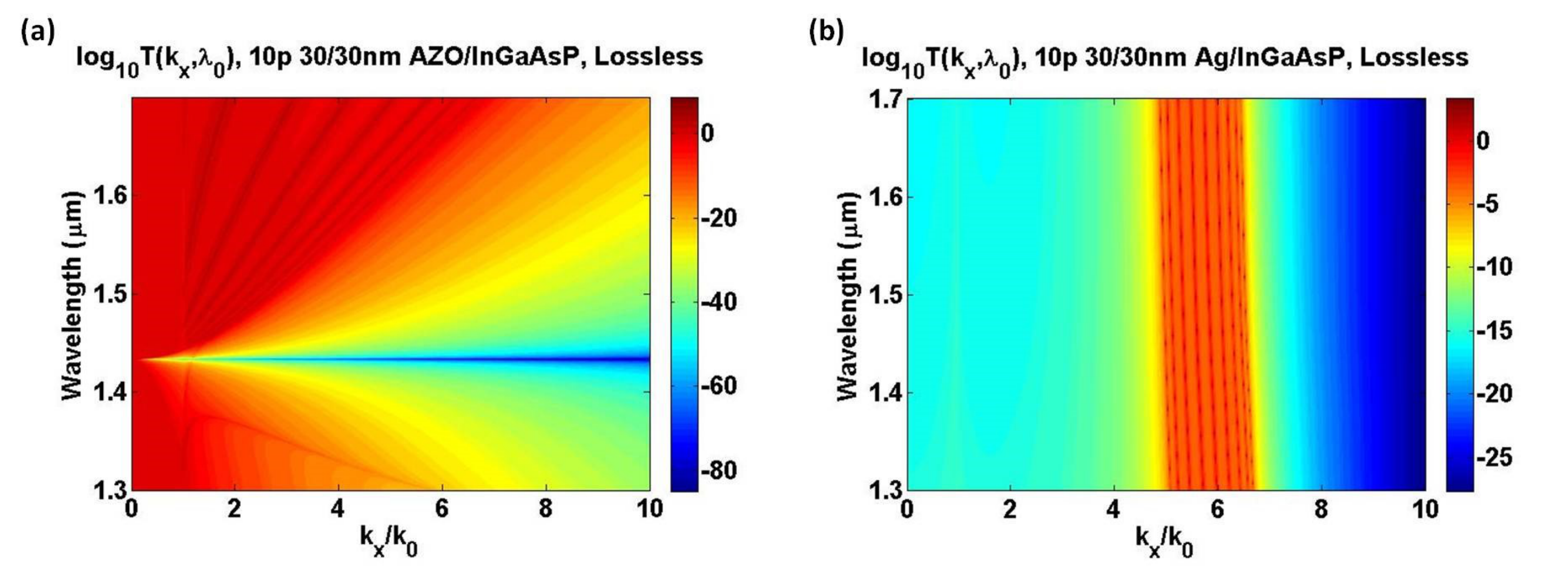}%
	\caption{\label{FigA1}Transmission contours for 10-period 30/30nm (a) AZO/InGaAsP and (b) Ag/InGaAsP systems with losses and gain omitted.  The greater-than-unity transmission in the absence of gain is an unphysical result that motivates the use of SMM. }
\end{figure}

Nevertheless, the TMM does correctly predict the location of transmission resonances in $k$-space.  Because it is computationally less demanding and simpler to implement than the SMM, it is useful for studying the dispersion of MD systems.  Figure \ref{FigA1} shows transmission contours with loss and gain omitted calculated via TMM for 10p 30/30nm MD systems.  Qualitatively, we observe that the AZO/InGaAsP system (Fig.\ref{FigA1}(a)) exhibits little dispersion ($d\lambda_0$/$dk_x$$\approx$0), while the Ag/InGaAsP system (Fig\ref{FigA1}(b)) is highly dispersive ($d\lambda_0$/$dk_x$$\gg$0).  This is a direct consequence of the fact that the plasma frequencies of AZO and Ag lie within and outside the NIR, respectively.  The greater-than-unity transmission in the absence of gain is a quantitatively unphysical result and motivates the use of the SMM for absolute transmission and relative transmission enhancement calculations.

\section{\label{App2}Discrepancy between SMM and FDTD}
\begin{figure}
	\includegraphics[width=9cm]{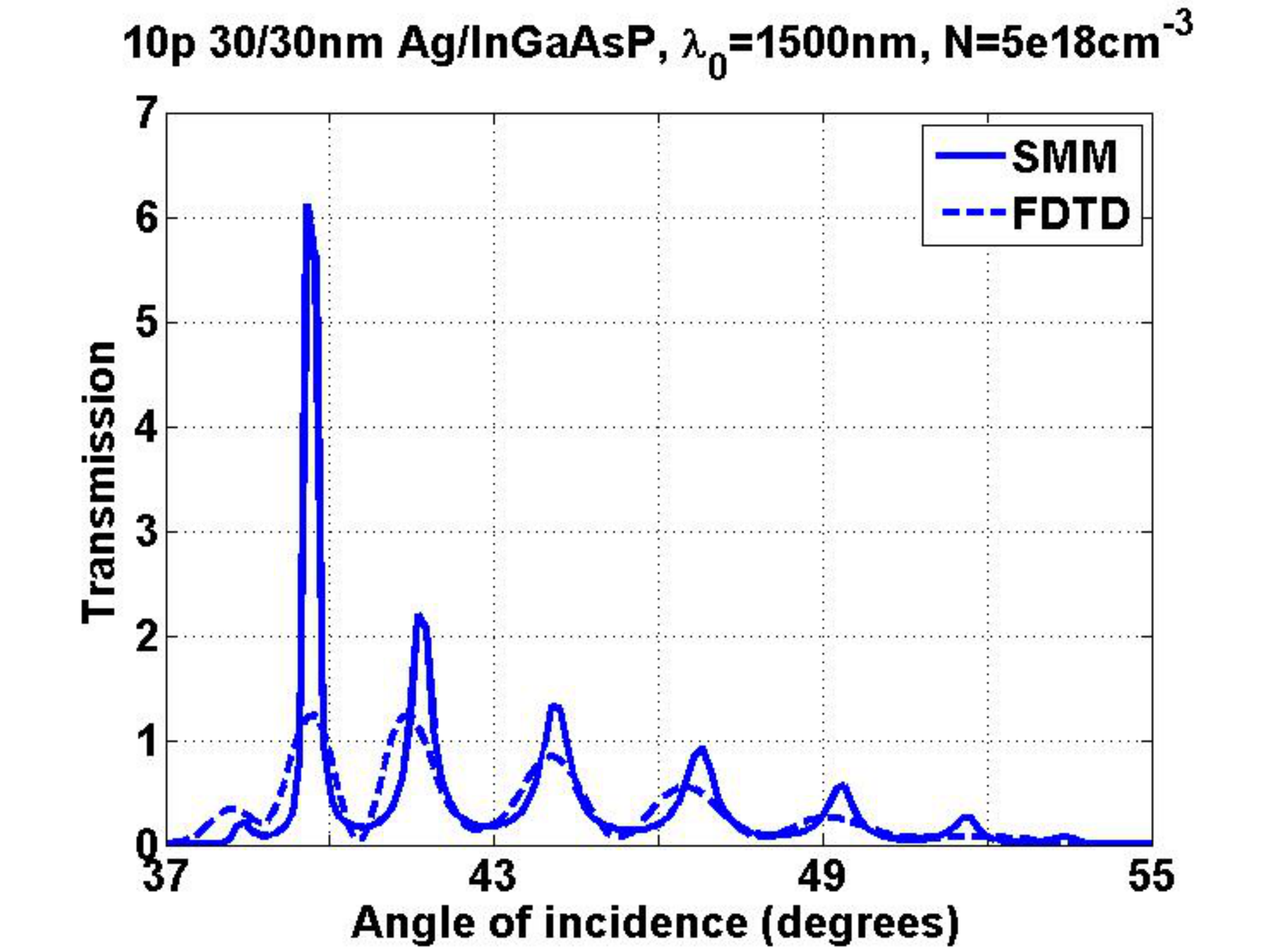}%
	\caption{\label{FigA2}Transmission through prism-coupled 10-period 30/30nm Ag/InGaAsP system under inversion at $\lambda_0$=1500nm. (SMM=scattering matrix method, FDTD=finite-difference time-domain)}
\end{figure}

The discrepancy in absolute transmission between the SMM and FDTD calculations is almost negligible at the high-$k$ end of the transmission window, but becomes significant at small values of $k_x$, in particular at the second angular resonance of Fig. \ref{FigA2}, which is calculated at an angular resolution of 7 samples per degree.  Potential reasons for this discrepancy rest in the finite-size mesh of the FDTD simulation and in the finite bandwidth of the FDTD source, compared to a monochromatic source in the SMM.  To check the former potential problem, we reduced the mesh size from the default minimum of 4nm to 0.1nm in both the direction normal and parallel to the layer interfaces.  We observed negligible change in the transmission upon reduction of the mesh size, indicating that a 4nm mesh size was adequate.  Given that the FDTD is exact once the finite spatial resolution is taken into account, we suspect that the SMM overestimates transmission, particularly at small values of $k_x$, due to its monochromatic nature.  In the SMM, all the energy gained by the input signal must be distributed across the range of simulated incident angles at a single frequency. In a system with optical losses but without gain, transmission is generally higher for smaller values of $k_x$ because the confinement to the lossy metal is less at smaller $k_x$.  Without being able to distribute the energy provided by the gain to other frequencies, the added energy in the SMM calculation with gain will increase the angular resonances.  For the FDTD however, energy gained is distributed across a range of frequencies.  In our FDTD simulations, the source has a finite bandwidth of 2.67x10$^{13}$Hz.  Arguably, the FDTD result is more physical because no source or transmitted signal is truly monochromatic.  Finally it should be emphasized that the discrepancy between the SMM and FDTD is relatively small compared to the discrepancy between the FDTD and TMM.  The TMM underestimates transmission by several orders of magnitude compared to the SMM and FDTD, and does this consistently over all angles of incidence.

\end{document}